\documentclass[twoside]{article}

\usepackage{PRIMEarxiv}

\usepackage[utf8]{inputenc} 
\usepackage[T1]{fontenc}    
\usepackage{hyperref}       
\usepackage{url}            
\usepackage{booktabs}       
\usepackage{amsfonts}       
\usepackage{nicefrac}       
\usepackage{microtype}      
\usepackage{lipsum}
\usepackage{fancyhdr}       
\usepackage{graphicx}       
\graphicspath{{media/}}     
\usepackage[super]{natbib}

\usepackage{graphicx}
\usepackage{amsmath,amssymb}
\usepackage{lineno}
\usepackage{multirow}
\usepackage{siunitx}
\usepackage{hyperref}
\hypersetup{
    colorlinks = true,
    allcolors = {black}
}
\usepackage{cleveref}

\crefname{equation}{Eq.}{Eqs.}
\Crefname{equation}{Equation}{Equations}

\crefname{table}{Table}{Tables}
\Crefname{table}{Table}{Tables}

\crefname{figure}{Fig.}{Figs.}
\Crefname{figure}{Figure}{Figures}

\crefname{section}{Sec.}{Secs.}
\Crefname{section}{Section}{Sections}

\crefname{chapter}{Chap.}{Chaps.}
\Crefname{section}{Section}{Sections}

\usepackage{tikz} 
\usetikzlibrary{arrows,scopes} 

\usepackage{geometry} 

\newcommand{\rc}{%
\resizebox{!}{1.25ex}{%
    \begin{tikzpicture}[>=round cap]
        \clip (0.09em,-0.05ex) rectangle (0.61em,0.81ex);
        \draw [line width=.11ex, <->, rounded corners=0.13ex] (0.1em,0.1ex) .. controls (0.24em,0.4ex) .. (0.35em,0.8ex) .. controls (0.29em,0.725ex) .. (0.25em,0.6ex) .. controls (0.7em,0.8ex) and (0.08em,-0.4ex) .. (0.55em,0.25ex);
    \end{tikzpicture}%
}%
}

\newcommand{\brc}{%
\resizebox{!}{1.3ex}{%
    \begin{tikzpicture}[>=round cap]
        \clip (0.085em,-0.1ex) rectangle (0.61em,0.875ex);
        \draw [line width=.2ex, <->, rounded corners=0.13ex] (0.1em,0.1ex) .. controls (0.24em,0.4ex) .. (0.35em,0.8ex) .. controls (0.29em,0.725ex) .. (0.25em,0.6ex) .. controls (0.7em,0.8ex) and (0.08em,-0.4ex) .. (0.55em,0.25ex);
    \end{tikzpicture}%
}%
}

\newcommand{\rce}{%
\resizebox{!}{1.00ex}{%
    \begin{tikzpicture}[>=round cap]
        \clip (0.09em,-0.05ex) rectangle (0.61em,0.81ex);
        \draw [line width=.09ex, <->, rounded corners=0.13ex] (0.05em,-0.1ex) .. controls (0.19em,0.2ex) .. (0.30em,0.6ex) .. controls (0.24em,0.525ex) .. (0.20em,0.4ex) .. controls (0.65em,0.6ex) and (0.03em,-0.4ex) .. (0.50em,0.05ex);
    \end{tikzpicture}%
}%
}

\title{Wavefield Correlation Imaging in Arbitrary Media with Inherent Aberration Correction}

\author{
  \href{https:\\www.scottschoenjr.com}{Scott Schoen~Jr} \\
  Harvard Medical School and Massachusetts General Hospital \\
  Boston, MA 02114 USA \\
  \& \\
  Georgia Institute of Technology \\
  Atlanta, GA 30332 USA \\
  \texttt{\href{mailto:scottschoenjr@gatech.edu}{scottschoenjr@gatech.edu}} \\
  \And
  Brian Lause, Rimon Tadross, \& Mike Washburn \\
  GE Healthcare \\
  Waukesha, WI 53188 \\
  \AND
  Marko Jokovljevic \& \href{https://curt.mgh.harvard.edu/}{Anthony E. Samir} \\
  Harvard Medical School and Massachusetts General Hospital \\
  Boston, MA 02114 USA \\
}

\begin{document}
\maketitle

\vspace{-1cm}
\begin{center}
\noindent\fbox{%
  \begin{minipage}{0.5\linewidth}
    \textit{The following article has been submitted to} Proceedings of Meetings on Acoustics. \textit{If accepted, it will be found at \href{https://pubs.aip.org/asa/poma}{pubs.aip.org/asa/poma}.}
  \end{minipage}%
}
\end{center}
\vspace{1cm}

\begin{abstract}
    Ultrasound~(US) imaging is an indispensable tool for diagnostic imaging, particularly given its cost, safety, and portability profiles compared to other modalities. 
    However, US is challenged in subjects with morphological heterogeneity (e.g., those with overweight or obesity), largely because conventional imaging algorithms do not account for such variation in the beamforming process.
    Specific knowledge of the these spatial variations enables supplemental corrections of these algorithms, but with added computational complexity.
    Wavefield correlation imaging~(WCI) enables efficient image formation in the spatial frequency domain that, in its canonical formulation, assumes a uniform medium.
    In this work, we present an extension of WCI to arbitrary known speed-of-sound distributions directly in the image formation process, and demonstrate its feasibility \textit{in silico}, \textit{in vitro}, and \textit{in vivo}.
    We report resolution improvements of over \SI{30}{\percent} and contrast improvements of order \SI{10}{\percent} over conventional WCI imaging.
    Together our results suggest heterogeneous WCI~(HWCI) may have high translational potential to improve the objective quality, and thus clinical utility, of ultrasound images.
\end{abstract}


\section{Introduction}
\label{intro}
Ultrasound (US) imaging, particularly for subjects with overweight or obesity, is challenged by aberration.\cite{brahee_body_2013}
Aberration manifests as artifacts in US images due to characteristics of the propagation of the incident or backscattered waves not accounted for by the imaging algorithm---most typically heterogeneity in the speed of sound in the medium.\cite{soulioti_deconstruction_2021}
Aberration compromises the diagnostic usefulness of US images; given the massive importance of the modality in diagnosing and monitoring disease, correcting aberration is of central importance to enhancing their value in clinical management.

We should emphasize at the outset that aberration correction in US imaging is a broad and active field, as degraded US imaging has implications for any approach that relies on an accurate mapping of the underlying physiology.
Myriad approaches have been proposed to correct aberration; a full review of these is out of scope for the present work, but a recent comprehensive review was recently presented by Ali et al.\cite{ali_aberration_2023}
Herein, we will consider the subset of approaches wherein the speed of sound within the medium is known (either \textit{a priori} or somehow inferred from the imaging data itself) and used to correct beamforming.

Most clinical US imaging employs the delay-and-sum algorithm~(DAS) to form images. 
In the conventional implementation of this approach, the time-of-flight from each source element in an ultrasound array to each position in the field of view, and then to each receiver element is computed, based on an assumed (typically uniform speed of sound) model of propagation.
Then, using these delays, the interpolated time-series signals from each are coherently summed to map the reflectivity of the medium at each point.
DAS is popular, as it is intuitive, it is fully parallelizable (enabling real-time imaging at several tens of hertz), and may be extended to account for nonuniform media by computing the spatially dependent delays (e.g., by integrating the slowness along each path, or by solving the eikonal equations for the known speed of sound field\cite{ali_distributed_2022}).

However, the latter case is significantly more computationally complex than homogeneous DAS, as the numerical integration or solving must be performed for each element in the array and every pixel in the image.
Given that the registration of the field to the array may change on a per-frame basis, such computational expense is not ideal.
The heterogeneous angular spectrum method~(HASM), a fast spatial frequency approach for propagating the acoustic field in arbitrary media, may be used to increase the speed of calculating these delays, though this approach requires robust 2D phase unwrapping, itself a nontrivial problem.\cite{schoen_fast_2023}

With the advent of software beamforming, other approaches to creating images from the backscattered echoes have become of interest.
For instance, short-lag spatial coherence beamforming\cite{lediju_short-lag_2011} creates a map of signal coherence, rather than reflection strength.
Recently, the approach of creating an image based on the correlation of the propagated transmitted and received wavefields, termed \emph{wavefield correlation imaging}~(WCI) was adapted from its original application in seismic imaging for use in medical ultrasound,\cite{roy_ultrasound_2016,theis_seismic_2020,ali_medical_2020} including formulations in the frequency domain\cite{ali_fourier-based_2021} and for curved arrays.\cite{ali_angular_2022}
In WCI, the known transmit signal is propagated forward, the received echoes are propagated backward, and the magnitude of the correlation between the two fields creates the image. 

The use of frequency domain correction schema has been proposed for WCI previously.
The simplest model is applicable to planar media only, and propagation is carried out in each constant-speed-of-sound layer sequentially.
While this approach was effective \textit{in vitro}, it cannot account for any lateral variation in the material properties.\cite{theis_seismic_2020,theis_seismic_2023}
Other works\cite{ali_distributed_2022,zhuang_simultaneous_2024} have employed the hybrid angular spectrum approach,\cite{vyas_ultrasound_2012} which accounts for more arbitrary variation in the speed of sound.
However, this approach requires knowledge of both the spatial and spatial frequency domain representations of the field at each step (HASM requires only the latter), and applies only phase corrections to the field. 
In this work, we propose the use of HASM in conjunction with WCI, termed heterogeneous wavefield correlation imaging~(HWCI), wherein the image is formed with the medium properties included directly in the computations, and thus the aberration correction is inherent.
After a brief overview of the theory, we characterize via objective quality metrics the improvement in image quality for simulated, phantom, and \textit{in vivo} data when HWCI is used to correct the aberration, to support the translational viability of the approach.

\section{Propagation Theory}
\subsection{Spatial Domain}
Forward, linear propagation of the field from a known source plane to arbitrary field points may be achieved via the Rayleigh integral.\cite{blackstock_fundamentals_2000,pierce_acoustics:_1989}
Without loss of generality, we can consider a time harmonic signal ($\propto \operatorname{exp}{-i\omega t}$), where $\omega$ is the angular frequency, and $t$ is time, such that $\partial/\partial t \rightarrow -i\omega$.
Taking the time dependence as implicit hereafter, then:
\begin{align}
    p(\boldsymbol{r}) 
    &=
    \frac{-i\omega \rho_{0}}{2\pi}
    \iint_{S}{%
        u_{z}(\boldsymbol{r}') \, g(\boldsymbol{r}, \boldsymbol{r}' ) \, \mathrm{d}x' \, \mathrm{d}y'
        },
    =
    \frac{-i\omega \rho_{0}}{2\pi}
    \iint_{S}{%
        u_{z}(\boldsymbol{r}') \, \frac{e^{i k \rce}}{\rc} \, \mathrm{d}x' \, \mathrm{d}y'
        },
    \label{eqn:RayleighIntegral}
\end{align}
where $\rho_{0}$ is density, $g$ is the free space Green's function, $u_{z}$ is the normal particle velocity at $z = 0$, $\boldsymbol{r}' = (x', y')$ are coordinates in the $z = 0$ plane, and
\begin{align}
    \rc 
    \equiv 
    \| \brc \|
    =
    \| \boldsymbol{r} - \boldsymbol{r}' \|
    =
    \sqrt{(x - x')^2 + (y - y')^2 + z^2}.
\end{align}
The direction of propagation may be reversed by conjugating \cref{eqn:RayleighIntegral}.

\subsection{Spatial Frequency Domain}
\Cref{eqn:RayleighIntegral} may be used to project and back-project analytically the transmitted and measured fields, respectively.
But, it is relatively computationally complex, and must be computed for each pixel within the image.
Equivalently, the field may be propagated via the angular spectrum method:\cite{williams_fourier_1999}
\begin{align}
    p(\boldsymbol{r})
    &=
    \mathcal{F}^{-1}_{k}\left[ P_{0}\,e^{i k_{z}z} \right]\,,
    \label{eqn:AngularSpectrum}
\end{align}
where $\mathcal{F}_{k}$ represents the 2D spatial Fourier transform, and $P_{0} = \mathcal{F}_{k}\left[ p(\boldsymbol{r}') \right]$ is the angular spectrum of the source condition (i.e., the pressure along the transducer surface), and $k_{z} = \sqrt{\omega^{2}/c_{0}^{2} - k_{x}^{2} - k_{y}^{2}}$.\footnote{%
Note that the Fourier transform of the Green's function is $\mathcal{F}_{k}[g(x,y,z)] = G(k_{x}, k_{y}, z) = e^{ik_{z}z}/2ik_{z}$, and that the definition of $\brc$ is equivalent to a convolution over $x$ and $y$ in \cref{eqn:RayleighIntegral}.
This is fundamentally a consequence of the convolution theorem.
}
The frequency domain version given by \cref{eqn:AngularSpectrum} is especially convenient as it enables computation of the field plane-by-plane, rather than point-by-point, significantly reducing the computational load, at the cost of only requiring execution of fast Fourier transforms.

\subsection{Heterogeneous Media}
In the case where the medium's speed of sound is not homogeneous, \cref{eqn:AngularSpectrum} is no longer valid.
By writing the heterogeneous Helmholtz equation as $\left[\nabla^{2} + \omega^{2}/c^{2}(\boldsymbol{r})\right]p = 0$, it may be shown\cite{schoen_heterogeneous_2020} that the projected field may be written as
\begin{align}
    p(x, y, \Delta z) 
    &= 
    \mathcal{F}_{k}^{-1}\left[
        P_{0}e^{ik_{z}\Delta z} 
        + 
        \frac{e^{ik_{z}\Delta z}}{2ik_{z}}\left( \Lambda * P_{0} \right)\Delta z
    \right]\,.
    \label{eqn:HeterogeneousAngularSpectrum}
\end{align}
In \cref{eqn:HeterogeneousAngularSpectrum}, $\Lambda = \mathcal{F}_{k}\left[ \left(  \omega^{2}/c_{0}^{2}- \omega^{2}/c^{2}(\boldsymbol{r} \right) \right]$ and $P_{0} = \mathcal{F}_{k}\left[ p(x, y, 0) \right]$.
Once the field at $z = \Delta z$ is computed, it may then used to compute the field at $2\Delta z$, and iteratively marched out to arbitrary depth $N\Delta z$.
Note that the spatial step need not be uniform, but it must be small relative to the wavelength, and the characteristic scale of the change of medium properties.\cite{schoen_acoustic_2020}

An alternative approach to \cref{eqn:HeterogeneousAngularSpectrum} is the split-step approach used by Refs.~\citenum{ali_distributed_2022} and \citenum{zhuang_simultaneous_2024}. 
Here, the first step is taken in the spatial frequency domain with a mean SoS for the depth, and then multiplied in the spatial domain by a laterally-varying correction term:
\begin{align}
    p(x, y, \Delta z)
    &=
    e^{i k_{\mathrm{res}}\Delta z}\mathcal{F}_{k}^{-1}\left[ 
        P_{0}\,e^{i\bar{k}_{z}\Delta z}
    \right]\,,
    \label{eqn:SplitStepAS}
\end{align}
where $\bar{k} = \omega\cdot \overline{c^{-1}}$ and $k_{\mathrm{res}} = \omega\cdot\left[c^{-1}(\boldsymbol{r}) - \overline{c^{-1}}\right]^{-1}$, and overbars indicate the mean value (note that $\overline{c^{-1}} \neq 1/\bar{c}$).
Comparing \cref{eqn:SplitStepAS} to \cref{eqn:HeterogeneousAngularSpectrum}, we note that HASM has two principle advantages: First, \cref{eqn:HeterogeneousAngularSpectrum} requires only knowledge of the angular spectrum $P(k_{x},k_{y},z)$ to compute $P(k_{x},k_{y},z + \Delta z)$, whereas \cref{eqn:SplitStepAS} requires both the angular spectrum of the field $P(k_{x},k_{y},z)$, as well as its spatial domain equivalent $p(x,y,z)$.
This requires an additional Fourier transform operation and consequently increases the computational expense.
Secondly, the second term within the brackets in \cref{eqn:HeterogeneousAngularSpectrum} enables a change of amplitude of each spectral component, whereas \cref{eqn:SplitStepAS} imparts only phase adjustments.


\section{Materials}
\label{sec:Methods}

\subsection{(Heterogeneous) Wavefield Correlation Imaging}
\begin{figure}
    \centering
    \includegraphics[width=\linewidth]{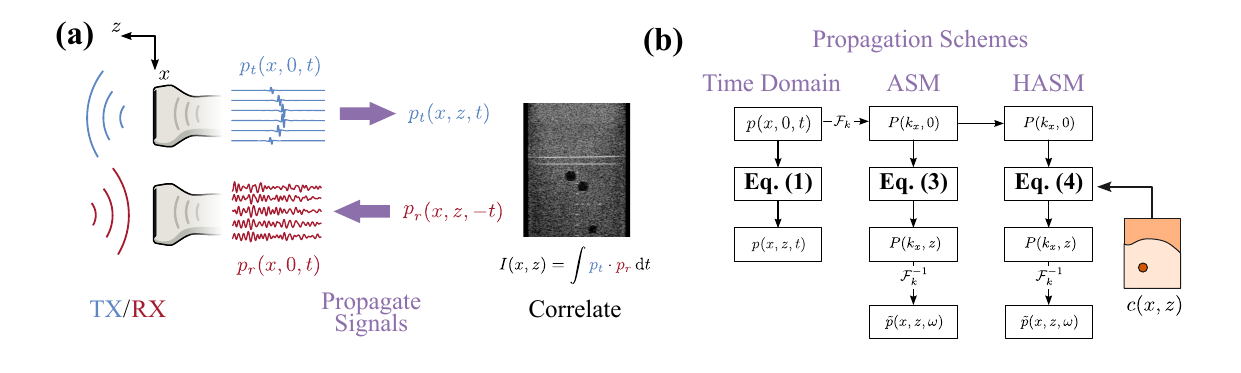}
    \caption{Wavefield Correlation Imaging Scheme for Ultrasound
    \textbf{(a)}~Transmitted (from known transmit delays) and received (measured) signals are propagated and backpropagated, respectively, to determine their wavefields $p(x,z,t)$. 
    The correlation in space (integrated over all times) defines the WCI image.
    \textbf{(b)}~The signals may be propagated in the time domain [by the Rayleigh integral, \cref{eqn:RayleighIntegral}], or in the spatial frequency domain.
    The angular spectrum method (ASM) assumes a uniform speed of sound, while HASM propagates the field using known properties $c(x,z)$.
    }
    \label{fig:WCIMethods}
\end{figure}

To compute the image maps, the propagated and back-propagated images were correlated via multiplication of their (time) frequency domain representations
\begin{align}
    I(x, z)
    &=
    \left\| \sum_{\omega}\left[ 
        p_{t}(x, z, \omega) \cdot p^{*}_{r}(x, z, \omega)
    \right] \right\|\,,
\end{align}
where the $t$ and $r$ subscripts indicate the transmitted and received fields, respectively, $(\cdot)^{*}$ indicates complex conjugation, and $\Omega$ is the bandwidth over which the image is formed.
The WCI concept is illustrated in \cref{fig:WCIMethods}, including the various propagation schemes.
Hereafter, ``WCI images'' will refer to those formed with $p_{t}(x, z, \omega)$ and $p_{r}(x, z, \omega)$ computed with \cref{eqn:AngularSpectrum}, while ``HWCI images'' will refer to those formed with the fields computed with \cref{eqn:HeterogeneousAngularSpectrum}.
In all cases, the bandwidth was selected to be \SI{90}{\percent} with respect to the transducer center frequency (e.g.,  $\Omega = [3.9, 10.3]$~MHz for the L11-5v simulations and experiments; see following sections).
Finally, data from the array were upsampled (linear interpolation) in the lateral direction by a factor of 2, apodized with a Tukey window, and zero padded such that the computational dimension size was 1.25 that representing the physical dimension.

\subsection{Simulations}
To generate realistic imaging conditions in changing media with known ground truth, a k-Wave\cite{treeby_k-wave_2010} simulation pipeline was created.
Here, the array was modeled as a Verasonics L11-5v probe (128 elements, pitch \SI{0.3}{mm}, \SI{7.1}{MHz} center frequency) with the \texttt{kArray} class.
The imaging sequence comprised 5 single-cycle plane wave emissions at \SI{5}{MHz}, tilted at \SI{0}{\degree}, $\pm$\SI{6}{\degree}, and $\pm$\SI{12}{\degree}.
The imaging medium was randomly generated using a modeled case of fat overlying liver, with an intervening layer of muscle.
Scattering was included by varying the local density by a factor of $5 \times 10^{-3}$ with respect to the medium's mean value.
One to four vessels were included with randomly determined transverse or lateral orientation.
The baseline material properties are given in \cref{tab:kWaveProperties}, and at the start of each simulation, values randomly chosen within \SI{3}{\percent} of the mean value were assigned to each region.
{\small %
\begin{table}
    \centering
    \begin{tabular}{ccccc}
        Medium & $c$ [m/s] & $\rho$ [kg/m\textsuperscript{3}] & $B/A$ & $\alpha_{0}$ [dB/cm/MHz{$^\gamma$}]  \\ \hline
        \textbf{Liver} & 1560 & 1070 & 6.8 & 0.06 \\
        \textbf{Muscle} & 1590 & 1050 & 7.4 & 0.35 \\
        \textbf{Fat} & 1440 & 900 & 10 & 0.22 \\
        \textbf{Blood} & 1580 & 1000 & 6.1 & 0.02 \\
    \end{tabular}
    \caption{Average Material Properties for k-Wave simulations.
    The constant frequency dependence exponent was set to $\gamma = 1.2$, and note the nonlinearity parameter is $B/A$ rather than the also-conventional quantity $1 + B/2A$.}
    \label{tab:kWaveProperties}
\end{table}
}
To enable evaluation of the imaging resolution, single-pixel ($\Delta x = \SI{200}{\micro\meter}$) inclusions with $c = \SI{2000}{m/s}$, $\rho = \SI{4000}{kg/m^{3}}$ were included at \SI{1}{cm} increments along the depth, and \SI{0.25}{cm} increments laterally.
The impedance mismatch was enforced largely through the density so that the CFL condition $\Delta x/c_{\mathrm{max}}\Delta t < 1$ would not require prohibitively small time or spatial discretizations.

\subsection{\textit{In Vitro} and \textit{In Vivo} Experiments}
\label{sec:MethodsInVitro}
A Verasonics Vantage 256 (Verasonics, Inc., Kirkland, WA USA) was programmed to image with the same conditions as the k-Wave simulations.
A linear L11-5v probe was used to interrogate a CIRS 040GSE (CIRS, Inc., Norfolk, VA USA).
As the speed of sound in the phantom was constant, heterogeneity was induced by overlaying gel pads (custom, CIRS, Inc.) with thickness \SI{2}{cm} and \SI{3}{\cm} and speeds of sound \SI{1400}{m/s} and \SI{1450}{m/s}, respectively.
Ultrasound coupling gel was used at the interface between the gel pad and phantom surface, as well as between the probe and gel pad.

All \textit{In Vivo} measurements were conducted in compliance with ethical practices and with approval from the Massachusetts General Hospital institutional review board~(IRB, protocol 2024P000359).
A Verasonics Vantage 256 was programmed to image with a linear probe (GE L2-9VN, 192 elements, center frequency \SI{6.0}{MHz}) to perform plane wave imaging with 21 plane waves with the same angular range as was used for the simulation (i.e., from \SI{-12}{\degree} to \SI{+12}{\degree}).
Transverse and longitudinal images of the carotid artery were obtained from a healthy 36-year-old male volunteer. 
To induce an aberration condition, a \SI{1}{cm} aberrating gel layer~(custom, CIRS, Inc.) with speed of sound \SI{1400}{m/s} was placed between the probe and neck of the subject.

\subsection{Evaluation Metrics}
To quantify the quality of the ultrasound images formed with WCI and HWCI, we employed several objective metrics.
The lateral full width at half maximum~(FWHM) of the point targets visualized in the simulations and phantom experiments serve as a measure of the resolution of the beamforming method.
Additionally the sharpness\cite{zhu_no-reference_2009} of the image, a unitless metric that captures the blur and noise present in an image, was used as a heuristic for the image quality.
Finally, to measure the contrast without respect to choices about the dynamic range, the generalized contrast-to-noise ratio~(gCNR)\cite{rodriguez-molares_generalized_2020} provides a normalized value from 0 to 1, where 1 denotes that there exists a cutoff intensity that divides all signal values and all background (or noise) values and enables perfect detectability of the feature.

\section{Results}
\label{sec:Results}

\subsection{Simulations}
\begin{figure}
    \centering
    \includegraphics[width=0.9\linewidth]{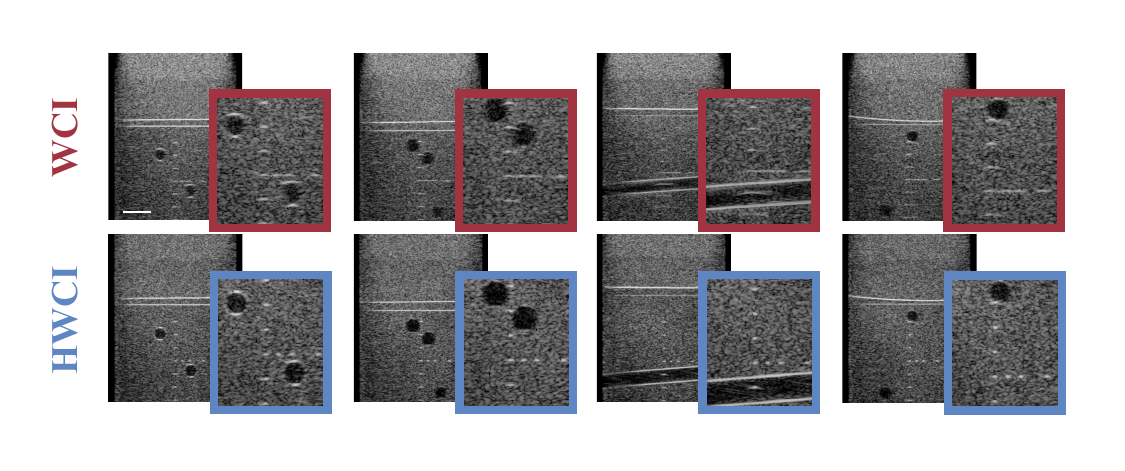}
    \caption{Representative \textit{in silico} images formed with WCI (top row, red) and HWCI (bottom row, blue).
    Insets show enlargements of pin targets and other features below the liver's interface with fat/muscle.
    Images are shown with \SI{40}{dB} dynamic range, and scale bar is \SI{1}{cm}.}
    \label{fig:ExampleSimImages}
\end{figure}

\begin{figure}
    \centering
    \includegraphics[width=0.7\linewidth]{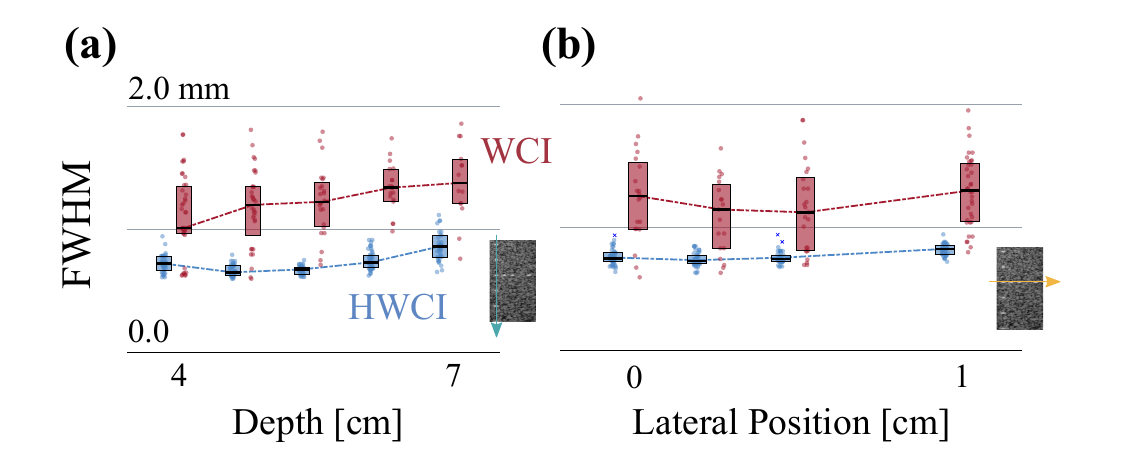}
    \caption{Lateral FWHM of pin targets in images formed with WCI (red) and HWCI (blue) from simulated k-Wave data.
    \text{(a)}~Variation in FWHM as a function of the depth (axial position) of the target.
    \text{(b)}~Variation in FWHM as a function of lateral position of the target.}
    \label{fig:PinFWHM}
\end{figure}

Overall, images formed with HWCI demonstrated reduced aberration effects: vessel boundaries were more prominent with improved boundary delineation, and pin targets showed reduced blurring (\cref{fig:ExampleSimImages}).
Specifically, the FWHM of point targets was significantly smaller in the HWCI case at all positions: for pins at the center of the image, the FWHM of the points was reduced by \SI{36.7\pm21.3}{\percent} [the \SI{95}{\percent} confidence interval~(CI) was (\SI{33.0}{\percent}, \SI{39.4}{\percent}) with $p < 10^{-5}$]; see \cref{fig:PinFWHM}(a).
For the pins at a fixed depth, the improvement was similarly significant, as seen in \cref{fig:PinFWHM}(b): \SI{31.4\pm20.1}{\percent} CI (\SI{28.6}{\percent}, \SI{34.2}{\percent}) and $p < 10^{-5}$.
HWCI also produced much more uniform point target widths. 
The coefficient of variation (i.e., the ratio between the standard deviation of the FWHMs at each target and their mean values) was \SI{28.1}{\percent} with WCI, compared with \SI{10.6}{\percent} for HWCI.

\subsection{\textit{In Vitro} and \textit{In Vivo} Experiments}
For the \textit{in vitro} reconstructions with HWCI, the speed of sound field was taken to be a planar layer (i.e., flat interface), and the nominal speeds of sound of each aberrating layer and the phantom were used in the image reconstruction.
The thickness of the layer was estimated from the image, as slight compression of the gel layers from their nominal thickness between the transducer and phantom was necessary to ensure good acoustic coupling. 
\begin{figure}[!htb]
    \centering    \includegraphics[width=0.8\linewidth]{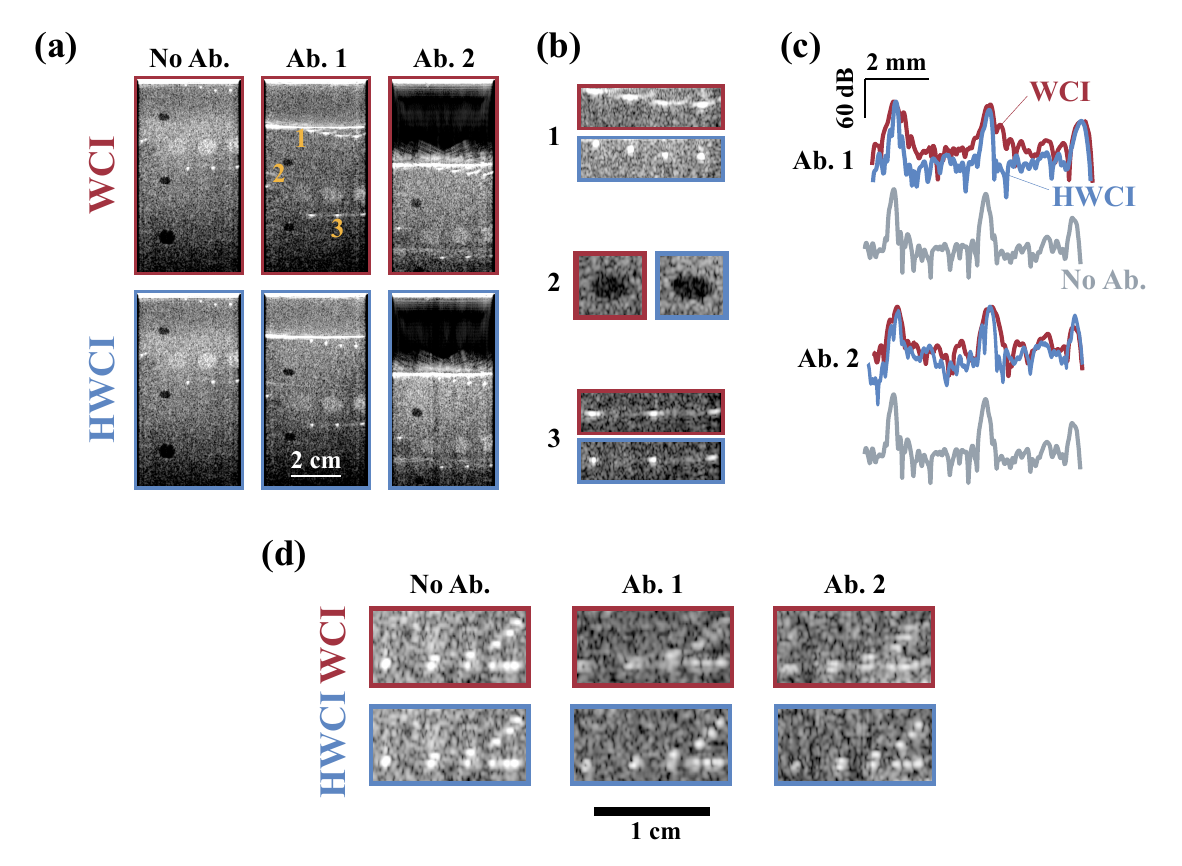}
    \caption{HWCI corrects aberration \textit{in vitro}.
    \textbf{(a)}~Phantom images formed with no aberrator, aberrator 1 (thickness \SI{2}{cm}, $c=\SI{1450}{m/s}$) and aberrator 2 (thickness \SI{3}{cm}, $c=\SI{1450}{m/s}$), formed with WCI (top row) and HWCI (bottom row).
    \textbf{(b)}~Enlarged views of the positions labeled in the central WCI image in (a) for WCI~(red) and HWCI~(blue).
    \textbf{(c)}~Intensity profiles for target 3 in (b), with each aberrator and compared to the no aberrator case (gray).
    \textbf{(d)}~Pin targets from another region of the phantom with each aberrator, formed using WCI~(top) and HWCI~(bottom). 
    }
    \label{fig:InVitroImages}
\end{figure}
\Cref{fig:InVitroImages} shows the images formed with WCI and HWCI for each aberrating layer.
Qualitatively, the HWCI images appear sharper and point targets retain features resembling those in the unaberrated images [\cref{fig:InVitroImages}(a--b)].
This effect was also clear in visualizations of point targets nominally at \SI{3}{cm} depth (i.e., at \SI{5.0}{cm} and \SI{6.0}{cm} depth for aberrators 1 and 2, respectively [\cref{fig:InVitroImages}(d)].

The improvement was reflected in objective metrics as well: HWCI improved the pin FWHM [measured at the pins labeled ``3'' in \cref{fig:InVitroImages}(a)] by \SI{34\pm17}{\percent} and \SI{40\pm23}{\percent} for aberrators 1 and 2, respectively, compared to WCI.
Similarly the gCNR [measured at the anechoic target labeled ``2'' in \cref{fig:InVitroImages}(a)] for HWCI compared to WCI improved by \SI{24.7}{\percent} and \SI{6.0}{\percent} for aberrators 1 and 2, respectively.
Finally, the sharpness of the images was higher in the HWCI image than in the WCI by \SI{14}{\percent} for aberrator 1 and \SI{9}{\percent} for aberrator 2.
\begin{table}[]
    \begin{center}
\begin{tabular}{ccccccc}
          & \multicolumn{2}{c}{\textbf{FWHM [m\textsuperscript{-4}]}}           & \multicolumn{2}{c}{\textbf{gCNR}} & \multicolumn{2}{c}{\textbf{Sharpness}} \\
Aberrator & WCI           & HWCI                        & WCI         & HWCI       & WCI           & HWCI          \\ \hline
Ab. 1      & 9.0 $\pm$ 2.4 & 6.2 $\pm$ 0.32 & 0.80        & 1.0        & 24.5          & 27.8          \\
Ab. 2      & 7.9 $\pm$ 2.0 & 4.5 $\pm$ 0.67 & 0.92        & 0.97       & 24.6          & 26.8         
\end{tabular}
    \caption{Objective image metrics for WCI and HWCI for each aberrator \textit{in vitro}. %
    FWHM are for profiles shown in \cref{fig:InVitroImages}(c), and gCNR were computed for anechoic target marked 2 in \cref{fig:InVitroImages}(a--b).}
    \label{tab:InVitroResults}
    \end{center}
\end{table}
These results are summarized in \cref{tab:InVitroResults}\,.

\begin{figure}[!htb]
    \centering
    \includegraphics[width=0.7\linewidth]{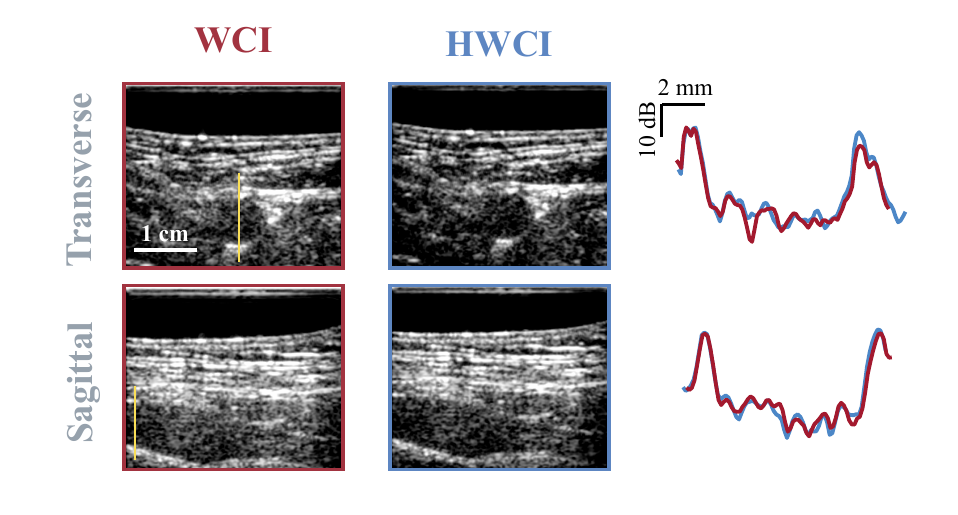}
    \caption{\textit{In vivo} images of the carotid artery in transverse (top row) and sagittal (bottom row) views, formed with WCI (blue) and HWCI (red).
    Right column shows the intensity profiles along the lines indicated in the leftmost images.
    Images are shown with \SI{30}{dB} of dynamic range}
    \label{fig:InVivoImages}
\end{figure}
Finally, for the \textit{in vivo} data, use of HWCI with the known parameters of aberrator 3 enabled a slight qualitative improvement, though little change was evinced by the quantitative metrics.
\Cref{fig:InVivoImages} shows the transverse (top) and sagittal (bottom) images of the carotid artery. 
While the artery appears slightly less cluttered in the HWCI compared to the WCI case, the gCNR changed by less than $10^{-3}$ between the two in both views.
Additionally, the line profiles taken show no significant difference between the corrected and uncorrected cases (\cref{fig:InVivoImages}, right).
The qualitative improvement was reflected in the sharpness metric, however, which improved by on the order of \SI{10}{\percent} when HWCI was used, compared to WCI (31.5 to 35.4 for sagittal, 29.3 to  32.4 for transverse).

\section{Discussion}
\label{sec:Discussion}
In this work, we have demonstrated that HASM improves resolution, sharpness, and contrast by incorporating the known material properties directly into the beamforming.
For realistic full wave simulations (\cref{fig:ExampleSimImages}), the HWCI images pin targets had FWHM on the order of \SI{35}{\percent} smaller than those in the conventional WCI images, and had significantly lower variability.
For \textit{in vitro} experiments with aberrating gel layers, a similar degree of improvement was observed (\cref{fig:InVitroImages} and \cref{tab:InVitroResults}).
Additionally for these cases, the image sharpness and gCNR were improved by on the order of \SI{10}{\percent} reflecting the qualitative improvement observed for the various targets within the phantom.
Finally, for a the \textit{in vivo} case, marginal improvements were computed in the image sharpness (though the contrast and gradients at vessel boundaries were approximately unchanged), with some apparent qualitative improvement.

There are a few limitations to the present study.
The central drawback to HWCI is the requirement that the speed of sound as a function of space is required, information typically not available for realistic imaging scenarios.
However, great progress has been made in recent years toward elucidating this information with ultrasound, by tracking spatial variations between transmits,\cite{sanabria_spatial_2018} local fitting of the received echos\cite{jakovljevic_local_2018}, optimization of coherence for layered media.\cite{ali_local_2022}
Additionally, recent years have seen a proliferation in the use of deep-learning~(DL) models to predict the speed of sound within the imaged medium,\cite{feigin_deep_2020,oh_learned_2021} including for its subsequent use to correct the beamforming.\cite{young_soundai_2022}
The ability of these  approaches to obtain SoS maps with accuracy on the order of a few percent suggests that they are sufficiently accurate to benefit the corrective abilities of HASM,\cite{schoen_experimental_2021} and thus HWCI.
Alternatively, one could use an empirical correction scheme for the backpropagation to obviate the need for a known SoS map.\cite{liu_correction_1994}

The requirement of a known speed-of-sound field also constrained the nature of the \textit{in vitro} and \text{in vitro} experiments possible in this work.
The results presented here involve only homogeneous layers (i.e., the aberrating gel layers with known speeds of sound). Thus the HWCI employed for the experiments are similar to previous results that employed planar layers and in principle require only sequential WCI to perform the correction.\cite{theis_seismic_2020}
However, the \textit{in silico} cases involved more varied media, including curved interfaces (\cref{fig:ExampleSimImages}). 
Provided an experimental arrangement with a known, laterally-varying speed of sound profile were available, these results suggest the correction would be effective.
We also investigated only plane wave imaging, mostly to reduce the number of k-Wave simulations required for the simulated cases.
While (H)WCI may also be used for focused transmits,\cite{ali_fourier-based_2021} the magnitude of improvement may differ from the results reported here.
Finally, we note that HASM includes only effects of varying speed of sound.
While they are expected to be less important, inclusion of nonlinear propagation effects and attenuation via an extended or higher-order marching scheme\cite{jing_k-space_2012,gu_modified_2020} may be of interest, e.g., for harmonic imaging.

\section{Conclusion}
\label{sec:Conclusion}
In this work, we have demonstrated that WCI may be augmented with a propagation scheme (HASM) that accounts efficiently for arbitrary medium heterogeneity and inherently corrects aberration during the image formation process (HWCI).
Compared to the conventional homogeneous WCI formulation, HWCI improved resolution on the order of \SI{35}{\percent} and improved contrast and sharpness by on the order of \SI{10}{\percent}, in simulated, \textit{in vitro}, and \textit{in vivo} ultrasound acquisitions.
Given its flexibility, straightforward implementation, and efficient frequency-domain implementation, HWCI has implications for improved imaging in cases challenged by aberration.

\section*{Acknowledgments}
Work supported by GE HealthCare

\bibliographystyle{unsrt}  
\bibliography{AberrationCorrection,GeneralAcoustics,SJS} 

\begin{thebibliography}{10}

\bibitem{brahee_body_2013}
Deborah~D. Brahee, Chinwe Ogedegbe, Cynthia Hassler, Themba Nyirenda, Vikki Hazelwood, Herman Morchel, Rita~S. Patel, and Joseph Feldman.
\newblock Body {Mass} {Index} and {Abdominal} {Ultrasound} {Image} {Quality}: {A} {Pilot} {Survey} of {Sonographers}.
\newblock {\em Journal of Diagnostic Medical Sonography}, 29(2):66--72, March 2013.
\newblock Publisher: SAGE Publications Inc STM.

\bibitem{soulioti_deconstruction_2021}
Danai Soulioti, Francisco Santibanez, and Gianmarco Pinton.
\newblock Deconstruction and reconstruction of image-degrading effects in the human abdomen: phase aberration, refraction, multiple reverberation, and trailing reverberation., September 2021.

\bibitem{ali_aberration_2023}
Rehman Ali, Thurston Brevett, Louise Zhuang, Hanna Bendjador, Anthony~S. Podkowa, Scott~S. Hsieh, Walter Simson, Sergio~J. Sanabria, Carl~D. Herickhoff, and Jeremy~J. Dahl.
\newblock Aberration correction in diagnostic ultrasound: {A} review of the prior field and current directions.
\newblock {\em Zeitschrift für Medizinische Physik}, 33(3):267--291, August 2023.

\bibitem{ali_distributed_2022}
Rehman Ali, Thurston Brevett, Dongwoon Hyun, Leandra~L. Brickson, and Jeremy~J. Dahl.
\newblock Distributed {Aberration} {Correction} {Techniques} {Based} on {Tomographic} {Sound} {Speed} {Estimates}.
\newblock {\em IEEE Transactions on Ultrasonics, Ferroelectrics, and Frequency Control}, 69(5):1714--1726, May 2022.

\bibitem{schoen_fast_2023}
Scott {Schoen Jr} and Anthony~E. Samir.
\newblock {Fast Spectral Approach for Delay Correction in Heterogeneous Media}.
\newblock In {\em Journal of the Acoustical Society of America}, Chicago, IL, 5 2023.

\bibitem{lediju_short-lag_2011}
Muyinatu~A. Lediju, Gregg~E. Trahey, Brett~C. Byram, and Jeremy~J. Dahl.
\newblock Short-{Lag} {Spatial} {Coherence} of {Backscattered} {Echoes}: {Imaging} {Characteristics}.
\newblock {\em IEEE transactions on ultrasonics, ferroelectrics, and frequency control}, 58(7):1377--1388, July 2011.

\bibitem{roy_ultrasound_2016}
O.~Roy, M.~a.~H. Zuberi, R.~G. Pratt, and N.~Duric.
\newblock Ultrasound breast imaging using frequency domain reverse time migration.
\newblock In {\em Medical {Imaging} 2016: {Ultrasonic} {Imaging} and {Tomography}}, volume 9790, pages 84--92. SPIE, April 2016.

\bibitem{theis_seismic_2020}
Daniela Theis and Ernesto Bonomi.
\newblock Seismic {Imaging} {Method} for {Medical} {Ultrasound} {Systems}.
\newblock {\em Physical Review Applied}, 14(3):034020, September 2020.
\newblock Publisher: American Physical Society.

\bibitem{ali_medical_2020}
Rehman Ali, Joseph Jennings, and Jeremy~J. Dahl.
\newblock Medical {Pulse}-{Echo} {Ultrasound} {Imaging} {Based} on the {Cross}-{Correlation} of {Transmitted} and {Backpropagated}-{Receive} {Wavefields}.
\newblock In {\em 2020 {IEEE} {International} {Ultrasonics} {Symposium} ({IUS})}, pages 1--4, September 2020.
\newblock ISSN: 1948-5727.

\bibitem{ali_fourier-based_2021}
Rehman Ali.
\newblock Fourier-based {Synthetic}-aperture {Imaging} for {Arbitrary} {Transmissions} by {Cross}-correlation of {Transmitted} and {Received} {Wave}-fields.
\newblock {\em Ultrasonic Imaging}, 43(5):282--294, September 2021.
\newblock Publisher: SAGE Publications Inc.

\bibitem{ali_angular_2022}
Rehman Ali and Jeremy Dahl.
\newblock Angular spectrum method for curvilinear arrays: {Theory} and application to {Fourier} beamforming.
\newblock {\em JASA Express Letters}, 2(5):052001, May 2022.

\bibitem{theis_seismic_2023}
Daniela Theis and Ernesto Bonomi.
\newblock Seismic imaging of medical ultrasound data: {Towards} in vivo applications.
\newblock {\em Europhysics Letters}, 142(5):52001, May 2023.
\newblock Publisher: EDP Sciences, IOP Publishing and Società Italiana di Fisica.

\bibitem{zhuang_simultaneous_2024}
Louise Zhuang, Thurston Brevett, Dongwoon Hyun, and Jeremy Dahl.
\newblock Simultaneous {Reverberation} {Noise} {Reduction} and {Aberration} {Correction} {Using} {Wavefield} {Correlation}.
\newblock In {\em 2024 {IEEE} {Ultrasonics}, {Ferroelectrics}, and {Frequency} {Control} {Joint} {Symposium} ({UFFC}-{JS})}, pages 1--5, September 2024.
\newblock ISSN: 2375-0448.

\bibitem{vyas_ultrasound_2012}
Urvi Vyas and Douglas Christensen.
\newblock Ultrasound beam simulations in inhomogeneous tissue geometries using the hybrid angular spectrum method.
\newblock {\em IEEE Transactions on Ultrasonics, Ferroelectrics, and Frequency Control}, 59(6):1093--1100, June 2012.
\newblock Conference Name: IEEE Transactions on Ultrasonics, Ferroelectrics, and Frequency Control.

\bibitem{blackstock_fundamentals_2000}
David~T. Blackstock.
\newblock {\em Fundamentals of {Physical} {Acoustics}}.
\newblock Wiley Interscience, New York, 2000.

\bibitem{pierce_acoustics:_1989}
Allan~D. Pierce.
\newblock {\em Acoustics: an {Introduction} to {Its} {Physical} {Principles} and {Applications}}.
\newblock Acoustical Society of America, Melville, NY, 1989.

\bibitem{williams_fourier_1999}
Earl~G. Williams.
\newblock {\em Fourier {Acoustics}: {Sound} {Radiation} and {Nearfield} {Acoustical} {Holography}}.
\newblock Academic Press, June 1999.
\newblock Google-Books-ID: vjfKLFBgMeIC.

\bibitem{schoen_heterogeneous_2020}
Scott Schoen and Costas~D. Arvanitis.
\newblock Heterogeneous {Angular} {Spectrum} {Method} for {Trans}-{Skull} {Imaging} and {Focusing}.
\newblock {\em IEEE Transactions on Medical Imaging}, 39(5):1605--1614, May 2020.
\newblock Conference Name: IEEE Transactions on Medical Imaging.

\bibitem{schoen_acoustic_2020}
Scott Schoen and Costas~D Arvanitis.
\newblock Acoustic source localization with the angular spectrum approach in continuously stratified media.
\newblock {\em The Journal of the Acoustical Society of America}, 148(4):EL333--EL339, October 2020.
\newblock Publisher: Acoustical Society of America.

\bibitem{treeby_k-wave_2010}
Bradley~E. Treeby and Benjamin~T. Cox.
\newblock k-{Wave}: {MATLAB} toolbox for the simulation and reconstruction of photoacoustic wave fields.
\newblock {\em Journal of Biomedical Optics}, 15(2):021314, March 2010.
\newblock Publisher: SPIE.

\bibitem{zhu_no-reference_2009}
Xiang Zhu and Peyman Milanfar.
\newblock A no-reference sharpness metric sensitive to blur and noise.
\newblock In {\em 2009 {International} {Workshop} on {Quality} of {Multimedia} {Experience}}, pages 64--69, July 2009.

\bibitem{rodriguez-molares_generalized_2020}
Alfonso Rodriguez-Molares, Ole Marius~Hoel Rindal, Jan D’hooge, Svein-Erik Måsøy, Andreas Austeng, Muyinatu~A. Lediju~Bell, and Hans Torp.
\newblock The {Generalized} {Contrast}-to-{Noise} {Ratio}: {A} {Formal} {Definition} for {Lesion} {Detectability}.
\newblock {\em IEEE Transactions on Ultrasonics, Ferroelectrics, and Frequency Control}, 67(4):745--759, April 2020.

\bibitem{sanabria_spatial_2018}
Sergio~J Sanabria, Ece Ozkan, Marga Rominger, and Orcun Goksel.
\newblock Spatial domain reconstruction for imaging speed-of-sound with pulse-echo ultrasound: simulation and in vivo study.
\newblock {\em Physics in Medicine \& Biology}, 63(21):215015, October 2018.
\newblock Publisher: IOP Publishing.

\bibitem{jakovljevic_local_2018}
Marko Jakovljevic, Scott Hsieh, Rehman Ali, Gustavo Chau Loo~Kung, Dongwoon Hyun, and Jeremy~J. Dahl.
\newblock Local speed of sound estimation in tissue using pulse-echo ultrasound: {Model}-based approach.
\newblock {\em The Journal of the Acoustical Society of America}, 144(1):254--266, July 2018.

\bibitem{ali_local_2022}
Rehman Ali, Arsenii~V. Telichko, Huaijun Wang, Uday~K. Sukumar, Jose~G. Vilches-Moure, Ramasamy Paulmurugan, and Jeremy~J. Dahl.
\newblock Local {Sound} {Speed} {Estimation} for {Pulse}-{Echo} {Ultrasound} in {Layered} {Media}.
\newblock {\em IEEE Transactions on Ultrasonics, Ferroelectrics, and Frequency Control}, 69(2):500--511, February 2022.
\newblock Conference Name: IEEE Transactions on Ultrasonics, Ferroelectrics, and Frequency Control.

\bibitem{feigin_deep_2020}
Micha Feigin, Daniel Freedman, and Brian~W. Anthony.
\newblock A {Deep} {Learning} {Framework} for {Single}-{Sided} {Sound} {Speed} {Inversion} in {Medical} {Ultrasound}.
\newblock {\em IEEE Transactions on Biomedical Engineering}, 67(4):1142--1151, April 2020.
\newblock Conference Name: IEEE Transactions on Biomedical Engineering.

\bibitem{oh_learned_2021}
SeokHwan Oh, Myeong-Gee Kim, YoungMin Kim, and Hyeon-Min Bae.
\newblock A {Learned} {Representation} {For} {Multi}-{Variable} {Ultrasonic} {Lesion} {Quantification}.
\newblock In {\em 2021 {IEEE} 18th {International} {Symposium} on {Biomedical} {Imaging} ({ISBI})}, pages 1177--1181, April 2021.
\newblock ISSN: 1945-8452.

\bibitem{young_soundai_2022}
James~R. Young, Scott Schoen, Viksit Kumar, Kai Thomenius, and Anthony~E. Samir.
\newblock {SoundAI}: {Improved} {Imaging} with {Learned} {Sound} {Speed} {Maps}.
\newblock In {\em 2022 {IEEE} {International} {Ultrasonics} {Symposium} ({IUS})}, pages 1--4, October 2022.
\newblock ISSN: 1948-5727.

\bibitem{schoen_experimental_2021}
Scott {Schoen Jr}, Pradosh~P. Dash, and Costas~D. {Arvanitis}.
\newblock {Experimental Demonstration of Trans-skull Volumetric Passive Acoustic Mapping with the Heterogeneous Angular Spectrum Approach}.
\newblock {\em IEEE Transactions on Ultrasonics, Ferroelectrics, and Frequency Control}, 69(2):534--542, 2022.

\bibitem{liu_correction_1994}
Dong‐Lai Liu and Robert~C. Waag.
\newblock Correction of ultrasonic wavefront distortion using backpropagation and a reference waveform method for time‐shift compensation.
\newblock {\em The Journal of the Acoustical Society of America}, 96(2):649--660, August 1994.

\bibitem{jing_k-space_2012}
Yun Jing, Tianren Wang, and Greg~T. Clement.
\newblock A k-{Space} {Method} for {Moderately} {Nonlinear} {Wave} {Propagation}.
\newblock {\em IEEE Transactions on Ultrasonics, Ferroelectrics, and Frequency Control}, 59(8):1664--1673, August 2012.
\newblock Conference Name: IEEE Transactions on Ultrasonics, Ferroelectrics, and Frequency Control.

\bibitem{gu_modified_2020}
Juanjuan Gu and Yun Jing.
\newblock A modified mixed domain method for modeling acoustic wave propagation in strongly heterogeneous media.
\newblock {\em The Journal of the Acoustical Society of America}, 147(6):4055--4068, June 2020.
\newblock Publisher: Acoustical Society of America.

\end{thebibliography}

\end{document}